\documentclass[11pt,showpacs,showkeys,footinbib,preprint,superscriptaddress,eqsecnum=true]{revtex4} 
\usepackage[utf8]{inputenc}
\usepackage[T1]{fontenc}
\usepackage{epsfig}
\usepackage{amsmath}
\usepackage{amsfonts}
\usepackage{amssymb}
\usepackage{color}
\usepackage{graphicx}
\usepackage{url,hyperref}
\usepackage{latexsym} 
\usepackage{longtable}
\usepackage{float}
\usepackage{epstopdf}
\begin{document}
	\title{Scattering states of position-dependent mass Schr\"{o}dinger equation with non central potential}
	\author{M. Chabab}
	\email{mchabab@uca.ma}
	\affiliation{High Energy Physics and Astrophysics Laboratory, Department of Physics, FSSM, Cadi Ayyad University P.O.B 2390, Marrakesh 40000, Morocco.}
	\author{A. El Batoul}
	\email{elbatoul.abdelwahed@edu.uca.ma}
	\affiliation{High Energy Physics and Astrophysics Laboratory, Department of Physics, FSSM, Cadi Ayyad University P.O.B 2390, Marrakesh 40000, Morocco.}
	\author{H. Hassanabadi}
	\email{h.hasanabadi@shahroodut.ac.ir}
	\affiliation{Department of Physics, University of Shahrood, Shahrood, Iran.}
	\author{M. Oulne}
	\email{oulne@uca.ma}
	\affiliation{High Energy Physics and Astrophysics Laboratory, Department of Physics, FSSM, Cadi Ayyad University P.O.B 2390, Marrakesh 40000, Morocco.}
	\author{S. Zare}
	\email{soroushzrg@gmail.com}
	\affiliation{Department of Basic Sciences, Islamic Azad University North Tehran Branch, Tehran, Iran.}
	\date{\today}
	\begin{abstract}
		In this paper, we study the time-independent Schr\"{o}dinger equation within  the formalism of position-dependent effective mass. For a generalized decomposition of the non-central effective potential, the deformed Schr\"{o}dinger equation can be easily solved analytically through separation of variables.  The energy eigenvalues and the normalization constant of the radial wave functions are obtained, as well as the scattering phase shifts.
	\end{abstract}	
\keywords{Schr\"{o}dinger equation, position-dependent mass, phase shift, scattering.}
	\pacs{03.65.Ge; 03.65.-w; 03.65.Pm}

	\maketitle
	\renewcommand{\theequation}{1.\arabic{equation}}
	\section{Introduction}
Conceptual meanings of the wave equations dependence on the exact solutions have been considered in \cite{b1,b2,b3,b4}. The derived physical quantities such as eigenfunctions and eigenvalues of these equations can be either  compared to the experimental or to the results obtained from other methods. The exact solutions of the wave equations can be used as criteria in other numerical and theoretical methods. The evolutions of non relativistic quantum particles are usually described by 
using the Schr\"{o}dinger equation while for the relativistic quantum particles, one has to deal with the correct equation of motion, such as Klein-Gordon or Dirac Equations, depending on the particle's spin. These equations have been investigated via different methods. Usually, the mass parameter in the above mentioned  wave equations has been considered to be a constant. Recently, an increasing interest has been devoted to solving quantum wave equations with position-dependent
mass. The Schr\"{o}dinger equation with position-dependent mass distribution was initially proposed by Von Roos \cite{b5}. In certain physical systems, the effective mass parameter should be position-dependent to be consistent with the experimental data \cite{b6}. In this context, the Schr\"{o}dinger equation with different phenomenological potentials and appropriate mass distributions has been investigated using various methods \cite{b7,b8,b9,b10}. For some molecular Hamiltonians, the energy spectra and eigenfunctions of the position-dependent mass particles have been derived \cite{b11}. According to Ref.\cite{b11}, particles with a mass are more likely to tunnel than ordinary ones. The use of this effective mass formalism has been considered for the dynamics of the electrons in inhomogeneous crystals for many years \cite{b12,b13}. It has been also applied in many different fields of physics, such as helium clusters\cite{b14}, semiconductors \cite{b15,b16,b17}, quantum dot\cite{b18}, quantum liquids\cite{b19}, atomic nuclei\cite{b20,b21,b22}.

In this work, we are going to consider the three dimensional time independent Schr\"{o}dinger equation within the effective mass formalism. This paper is organized as follows. In section II, after some preliminaries, the separation of variables is carried out for the deformed Schr\"{o}dinger equation  with non-central potential in spherical coordinates.
 In section III, we introduce a new non-central potential and we investigate the scattering states solutions as well as the phase shifts under this effective potential . Finally, Sec. IV is devoted to the conclusion.
 
\section{Variable Separation of Hamiltonian Considering Non-Central Potential in Position-Dependent Mass Formalism }

\numberwithin{equation}{section}

Theoretical background of the position-dependent effective mass formalism (PDEMF) was recently has been considered  in Refs \cite{b23,b24}. In the PDEMF, for Schr\"{o}dinger equation, the mass operator $m(x)$ and momentum operator $\vec{p}=-i\hbar\vec{\nabla}$ no longer commute. Therefore, several ways exist for generalizing the usual form of the kinetic energy operator $\vec{p}^2/2m_0$,  and consequently the Hamiltonian, in order to obtain a Hermitian operator to describe the quantum state of a physical system that is not trivial. In order to avert any specific choices, one can use the general form of the Hamiltonian originally proposed by Von Roos \cite{b25}.
In Ref. \cite{b26}  by choosing the position-dependent mass $m(\vec{r})=\frac{m_0}{f(\vec{r})^2}$ where $m_0$ is a constant mass and $f(r)$ represents a deforming function, the authors obtain a new form of Hamiltonian, and in the special case, this Hamiltonian reduces to the very common BenDaniel-Duke form \cite{b27}.
In the spherical coordinates $\vec{r}=\{r=||\vec{r}||,\theta,\varphi\}$ with $f(\vec{r})=f(r)$, follow separation of variables is customary used to obtain the wave function as  
$\psi(r,\theta,\varphi)=\frac{1}{r}\frac{U(r)}{f(r)}Y_{(\ell)}^{(\Lambda)}(\theta,\varphi)$, with $Y_{(\ell)}^{(\Lambda)}(\theta,\varphi)=\Theta(\theta)\Phi(\varphi)$, for a  potential taking the form \cite{b26},
\begin{equation}
V(r,\theta,\varphi)=V_1(r)+\frac{f(r)^2}{r^2}V_2(\theta)+\frac{f(r)^2}{r^2sin^2(\theta)}V_3(\varphi)
\label{E9}
\end{equation}
where $V_1(r)$, $V_2(\theta)$ and $V_3(\varphi)$ are arbitrary functions depending on specific arguments.\\
Then, the position dependent mass Schr\"{o}dinger equation  with the non central potential defined in (\ref{E9}) can be transformed into a separate system in all three coordinates \cite{b26}:
\begin{eqnarray}
\label{E10}
\bigg[\frac{d^2}{dr^2}+\frac{2m_0}{\hbar^2}\left(\frac{E-V_1(r)}{f(r)^2}\right)-\frac{L^2}{r^2}- \bar{F}(r,\lambda,\delta) \bigg]U(r)=0, \ r\in[0,\infty]
\end{eqnarray}
where $\bar{F}(r,\lambda,\delta) =\frac{(2-\delta-\lambda)}{f(r)}\left(\frac{f''(r)}{2}+\frac{f'(r)}{r}\right) \left(\left(\frac{1}{2}-\delta\right)\left(\frac{1}{2}-\lambda\right)-\frac{1}{4}\right) \left(\frac{f'(r)}{f(r)}\right)^2$
\begin{align}
&\left[\frac{d^2}{d\theta^2}+cot(\theta)\frac{d}{d\theta}+L^2-\frac{\Lambda^2}{sin^2(\theta)}-\frac{2m_0}{\hbar^2}V_2(\theta)\right]\Theta(\theta)=0, \theta\in[0,\pi]
\label{E11}\\
&\left[\frac{d^2}{d\varphi^2}-\frac{2m_0}{\hbar^2}V_3(\varphi)+\Lambda^2\right]\Phi(\varphi)=0,\  \varphi\in[0,2\pi]
\label{E12}
\end{align}
where $\Lambda^2$ and $L^2=\ell(\ell+1)$ are real and dimensionless separation constants. The components of the wavefunction are also constrained to satisfy the boundary conditions: $U(0)=U(\infty)=0$ for the bound states, or $U(0)=0$ for the continuous states, $ \Phi(\varphi)=\Phi(\varphi+2\pi)$, while $\Theta(0)$ and $\Theta(\pi)$ are finite.

\section{ Scattering state solutions and  Phase Shifts}
In this section, we consider a particle influenced by a new non central potential, dubbed  the double ring-shaped polynomial field potential, obtained from Eq.(\ref{E9}) with:

\begin{align}
	& V_1(r)=a+b\cdot r+c\cdot r^2, \label{E14} \\
	& V_2(\theta)=\left( \frac{B}{sin^2(\theta)}+\frac{A(A-1)}{cos^2(\theta)}\right)
	\label{E15}
    \\
	& V_3(\varphi)=\left( \frac{\alpha^2D(D-1)}{sin^2(\alpha\varphi)}+\frac{\alpha^2C(C-1)}{cos^2(\alpha\varphi)}\right)
	\label{E16}
\end{align}

where the parameters are chosen as $ A,C,D > 1; a,b, B\geq0; c=\frac{1}{2}m_0\omega^2$,  $\alpha=1,2,3,\cdots$. 
When $a=b=0$ and $D=C=1$, the potential reduces to a double-ring-shaped oscillator potential. Also, when $a=b=B=0$ and $A=C = D = 1$, it reduces to a spherical oscillator potential, considered as one of the most important models in classical and quantum physics.

In the subsequent subsection, we are going to study the  scattering states of the Schr\"{o}dinger equation with the double ring-shaped polynomial field potential in the spherical coordinates.
\subsection{Exact solutions of the first angular equation}
 We start our investigation with the angular $\varphi$ part  of the Schr\"{o}dinger equation. 
 After introducing the shape form of the potential shown in Eq. (\ref{E16}) into Eq.(\ref{E12})  we get,
 \begin{equation}
 \left[\frac{d^2}{d\varphi^2}+\Lambda^2-\frac{2m_0}{\hbar^2}\left(\frac{\alpha^2D(D-1)}{sin^2(\alpha\varphi)}+\frac{\alpha^2C(C-1)}{cos^2(\alpha\varphi)}\right)\right]\Phi(\varphi)=0
 \label{E17}
 \end{equation}
 By defining a new variable $x=\sin(\alpha\varphi)^2$, this equation  transforms into,

 \begin{equation}
 \frac{d^2\Phi(x)}{dx^2}+\frac{\frac{1}{2}-x}{x(1-x)}\frac{d\Phi(x)}{dx}+\frac{(-\xi_1^2x^2+\xi_2^2x-\xi_3^2)}{x^2(1-x)^2}\Phi(x)=0
 \label{E18}
 \end{equation}

 with

 \begin{align}
 & \xi_1^2=\frac{\Lambda^2}{4\alpha^2},\label{E19}\\ 
 & \xi_2^2=\frac{m_0}{2\hbar^2}\left(D(D-1)-C(C-1)\right)+\frac{\Lambda^2}{4\alpha^2},\label{E20}\\
 & \xi_3^2=\frac{m_0}{2\hbar^2}D(D-1)
 \label{E21}
 \end{align}

 According to the Nikiforov-Uvarov procedure, the resulting energy eigenvalues are,

 \begin{eqnarray}
 	n_{\varphi}^2+\left(2n_{\varphi}+1\right)\left( \left(\frac{1}{16}+\left(\xi_1^2+\xi_3^2-\xi_2^2\right)\right)^{\frac{1}{2}}+\left(\frac{1}{16}+\xi_3^2\right)^{\frac{1}{2}}+\frac{1}{2}\right)+\nonumber \\
 	2\left(\frac{1}{16}+\left(\xi_1^2+\xi_3^2-\xi_2^2\right)\right)^{\frac{1}{2}}\left(\frac{1}{16}+\xi_3^2\right)^{\frac{1}{2}}+\left(2\xi_3^2-\xi_2^2-\frac{1}{8}\right)=0
 \label{E22}
 \end{eqnarray} 
Substituting $\xi_1$, $\xi_2$ and $\xi_3$ by their expressions given in Eq. (\ref{E19}), Eq. (\ref{E20}) and Eq. (\ref{E21}) respectively, we finally derive the exact formula of $\Lambda$
 \begin{equation}
 \Lambda=\pm\alpha\left(\frac{\sqrt{1+\frac{8m_0C(C-1)}{\hbar^2}}}{2}+\frac{\sqrt{1+\frac{8m_0D(D-1)}{\hbar^2}}}{2}+2n_{\varphi}+1\right),\ n_{\varphi}=0,1,2,\cdots
 \label{E23}
 \end{equation}
 which exactly reproduce the result reported in \cite{b26}.
 
 The corresponding eigenfunctions  of Eq. (\ref{E18}) read as,

 \begin{equation}
 \Phi(x)=x^{\frac{1}{2}+\left(\frac{1}{16}+\xi_3^2\right)^{\frac{1}{2}}}\left(1-x\right)^{\frac{1}{4}+\left(\frac{1}{16}+\xi_1^2+\xi_3^2-\xi_2^2\right)^{\frac{1}{2}}}
 P_{n_{\varphi}}^{\left(  \left(\frac{1}{4}+4\xi_3^2\right)^{\frac{1}{2}},  \left(\frac{1}{4}+4\left(\xi_1^2+\xi_2^2-\xi_3^2\right) \right)^{\frac{1}{2}} \right)}\left(1-2x\right)
 \label{E24}
 \end{equation}

where $P_{n}^{(a,b)}(z)$ is the generalised Jacobi functions.

 \subsection{Exact solutions of the second angular equation}
 The substitution of the potential (\ref{E14}) into Eq. (\ref{E10}) leads to the following differential equation :
 \begin{equation}
 \left[\frac{d^2}{d\theta^2}+cot(\theta)\frac{d}{d\theta}+L^2-\frac{\Lambda^2}{sin^2(\theta)}-\frac{2m_0}{\hbar^2}\left\lbrace \frac{B}{sin^2(\theta)}+\frac{A(A-1)}{cos^2(\theta)}\right\rbrace \right]\Theta(\theta)=0
 \label{E25}
 \end{equation}
 To solve this equation, we also introduce the transformation $z=\cos(\theta)^2$, so we obtain,

  \begin{equation}
  \frac{d^2\Theta(z)}{dz^2}+\frac{1-\frac{3}{2}z}{z(1-z)}\frac{d\Theta(z)}{dz}+\frac{\left(-\chi_1^2z^2+\chi_2^2z-\chi_3^2\right)}{z^2(1-z)^2}\Theta(z)=0,
  \label{E26}
  \end{equation}
 with
 \begin{align}
 &\chi_1^2=\frac{L^2}{4}=\frac{\ell(\ell+1)}{4}
 \label{E27}
 \\
 &\chi_2^2=\frac{m_0}{2\hbar^2}\left(B-A\left(A-1\right)\right)+\frac{1}{4}\left(L^2+\Lambda^2\right)
 \label{E28}
 \\
 &\chi_3^2=\frac{m_0}{2\hbar^2}B+\frac{\Lambda^2}{4}
 \label{E29}
 \end{align}
 Like Eq. (\ref{E18}), the eigenfunctions of Eq. (\ref{E26}) are the generalised Jacobi functions,

 \begin{equation}
 \Theta(z)=z^{\chi_3}\left(1-z\right)^{\frac{1}{4}+\left(\frac{1}{16}+\chi_1^2+\chi_3^2-\chi_2^2\right)^{\frac{1}{2}}}P_{n_{\theta}}^{\left( 2\chi_3 ,  \left(\frac{1}{4}+4\left(\chi_1^2+\chi_2^2-\chi_3^2\right) \right)^{\frac{1}{2}} \right)}\left(1-2z\right)
 \label{E30}
 \end{equation}

 and the corresponding eigenvalues are solutions of the equation,

 \begin{eqnarray}
 &\frac{1}{2}n_{\theta}+n_{\theta}^2+\left(2n_{\theta}+1\right)\left(\left(\frac{1}{16}+\chi_1^2+\chi_3^2-\chi_2^2\right)^{\frac{1}{2}}+\chi_3+\frac{1}{4}\right)+2\chi_3\left(\frac{1}{16}+\chi_1^2+\chi_3^2-\chi_2^2\right)^{\frac{1}{2}}\nonumber \\&+2\chi_3^2-\chi_2^2=0
 \label{E31}
 \end{eqnarray}

 Substituting $\chi_1$, $\chi_2$ and  $\chi_2$ by their expressions shown in Eq. (\ref{E27}), Eq. (\ref{E28}) and Eq. (\ref{E29}) respectively, we finally obtain the full expression of $\ell$:
 \begin{equation}
 \ell=\frac{1+\sqrt{1+\frac{8m_0A(A-1)}{\hbar^2}}}{2}+\sqrt{\Lambda^2+\frac{2m_0B}{\hbar^2}}+2n_{\theta},\ n_{\theta}=0,1,2,\cdots
 \label{E32}
 \end{equation}
 which again coincides with the formula derived in \cite{b26} using the asymptotic iteration method.
 
 \subsection{Scattering Phase Shifts}
In order to study the  scattering state and phase shifts in the problem of position-dependent mass Schr\"{o}dinger equation with the double ring shaped polynomial field potential given by Eq.(\ref{E13}), we must define the deformation function $f(r)$. Thus, in this section, we use a simple linear representation \cite{b26},

\begin{equation}
	f(r)=1+f_0 r,\  
	\label{E33}
\end{equation}  

By substituting into  Eq. (\ref{E10}) and by using the potential (\ref{E14}), the confluent forms of Heun's differential equation show up:
\begin{equation}
 \left[\frac{d^2}{dr^2}+\frac{2m_0}{\hbar^2(1+f_0r)^2}\left(E-a-cr^2-br-\frac{\hbar^2}{2m_0}P\right)-\frac{Q}{r(1+f_0r)}-\frac{L^2}{r^2}\right]U(r)=0
 \label{HeunE}
\end{equation}
with
\begin{eqnarray}
P=\left[\left(\frac{1}{2}-\lambda\right)\left(\frac{1}{2}-\delta\right)-\frac{1}{4}\right]f_0,\ Q=\left[2-\delta-\lambda\right]f_0
\end{eqnarray}

This equation is not easy to solve, however, because we are dealing with scattering states,  we can safely replace $1+f_0r$ by $f_0r$ in Eq. (\ref{HeunE}) with $f_0 > 0$. Consequently, the above Heun's  equation simplifies to the following differential equation,  

 \begin{equation}
  \left[\frac{d^2}{dr^2}+\left(-\frac{\ell'\left(\ell'+1\right)}{r^2}+\bar{K}^2+\frac{2\bar{\Lambda}}{r}\right)\right]U_{n,\ell'}(r)=0,
  \label{E34}
 \end{equation}

 with 
 
 \begin{align}
& \ell'\left(\ell'+1\right)=\ell(\ell+1)-\frac{2m_0}{\hbar^2}\left(\frac{E-a}{f_0^2}\right)+\left(\frac{1}{2}-\delta\right)\left(\frac{1}{2}-\lambda\right)-(\lambda+\delta)+\frac{7}{4} 
\label{E35}
\\
& \bar{\Lambda}=-\frac{m_0}{\hbar^2}\frac{b}{f_0^2}
\label{E36}
 \\
& \bar{K}^2= -\frac{m_0}{\hbar^2}\frac{2c}{f_0^2}
\label{E37}
 \end{align}
 
Notice here that the $\ell'$ parameter plays the role of the orbital angular momentum in problems with  spherical central  potentials.
 Having in mind the boundary conditions of the scattering states, i.e. $U_{n,\ell'}(r=0)=0$, we use the following ansatz for the asymptotic behavior of the wave function at the origin:
 \begin{equation}
U_{n,\ell'}(r)=A\cdot (\bar{K}r)^{\ell'+1}e^{i\bar{K}r}\xi_{n,\ell'}(r)
\label{E38}
 \end{equation}
 Insertion of Eq. (\ref{E38}) into Eq. (\ref{E34}) results in
 \begin{equation}
 \left[r\frac{d^2}{dr^2}+\left(2\ell'+2i\bar{K}r+2\right)\frac{d}{dr}+\left(2\bar{\Lambda}+2i\bar{K}\left(\ell'+1\right)\right)\right]\xi_{n,\ell'}(r)=0
 \label{E39}
 \end{equation}
 If in addition we introduce a new variable $s =- 2i\bar{K}r$, then Eq. (\ref{E39}) can be alternatively written as
 \begin{equation}
 \left[s\frac{d^2}{ds^2}+\left(2\ell'+2-s\right)\frac{d}{ds}+\left(\ell'+1-i\frac{\bar{\Lambda}}{\bar{K}}\right)\right]\xi_{n,\ell'}(s)=0,
 \label{E40}
 \end{equation}
 with $s=|s|e^{-i\frac{\pi}{2}}$. Equation (\ref{E40}) is just the confluent Hypergeometric equation. General form of confluent Hypergeometric can be written as 
 \begin{equation*}
 z\frac{{{d^2}w}}{{d{z^2}}} + (b - z)\frac{{dw}}{{dz}} - aw = 0,
 \end{equation*}
  where $a$ and $b$ are constant. The Solution of above equation can be written with the aid of Kummer's functions as 
 \begin{equation*}
 M\left( {a,b,z} \right) = \sum\limits_{n = 0}^\infty  {\left( {\frac{{{a^{\left( n \right)}}{z^n}}}{{{b^{\left( n \right)}}n!}}} \right) = {}_1{F_1}} \left( {a,b,z} \right),
 \end{equation*}
 comparing \eqref{E40} with the general form of confluent Hypergeometric differential equation results in
  , as $r\rightarrow 0$
  \begin{equation}
\xi_{n,\ell'}(s)={}_1F_1\left(\ell'+1-i\frac{\bar{\Lambda}}{\bar{K}},2\ell'+2,-2i\bar{K}r\right).
\label{E41}
  \end{equation}
 Therefore, our analytical expression of the radial wave function for the scattering states is obtained by substituting Eq.\eqref{E41} into Eq.\eqref{E38} :
  \begin{equation}
  U_{n,\ell'}(r)=A\cdot (\bar{K}r)^{\ell'+1}e^{i\bar{K}r}{}_1F_1\left(\ell'+1-i\frac{\bar{\Lambda}}{\bar{K}},2\ell'+2,-2i\bar{K}r\right).
  \label{E42}
  \end{equation}
  Now, we want to obtain the asymptotic behavior of the wave function for $r\rightarrow 0$, then calculate the normalization constant and the phase shifts. To this end, we use the transformation formulae for confluent Hypergeometric function when $s\rightarrow \infty$:
  \begin{equation}
  {}_1F_1\left(\alpha,\gamma,s\right)=\frac{\Gamma(\gamma)}{\Gamma(\alpha)}e^ss^{\alpha-\gamma}+\frac{\Gamma(\gamma)}{\Gamma(\gamma-\alpha)}e^{\pm i\pi\alpha}s^{-\alpha}
  \label{E43}
  \end{equation}
  where $"+"$ and  $"-"$ correspond to  $arg(s)\in]-\pi/2,3\pi/2[$ and  $arg(s)\in]-3\pi/2,\pi/2[$ respectively. 
 By subtituting $s=|s|e^{-i\frac{\pi}{2}}$, Eq. (\ref{E43}) is re-expressed as
  \begin{equation}
  {}_1F_1\left(\alpha,\gamma,s\right)=\frac{\Gamma(\gamma)}{\Gamma(\alpha)}e^s|s|^{\alpha-\gamma}e^{-i(\alpha-\gamma)\pi/2}+\frac{\Gamma(\gamma)}{\Gamma(\gamma-\alpha)}e^{ i\pi\alpha/2}|s|^{-\alpha}
  \label{E44}
  \end{equation}
 from which we obtain 
  \begin{eqnarray}
  &{}_1F_1\left(\ell'+1-i\frac{\bar{\Lambda}}{\bar{K}},2\ell'+2,-2i\bar{K}r\right)\nonumber  = \frac{\Gamma(2\ell'+2)}{\Gamma(\ell'+1-i\frac{\bar{\Lambda}}{\bar{K}})}e^{-2i\bar{K}}\left(2\bar{K}r\right)^{-\left(\ell'+1+i\bar{\Lambda}/\bar{K}\right)} 
   e^{i\left(\ell'+1+i\bar{\Lambda}/\bar{K}\right)\pi/2} \nonumber\\
   &+\frac{\Gamma(2\ell'+2)}{\Gamma(\ell'+1+i\frac{\bar{\Lambda}}{\bar{K}})}\left(2\bar{K}r\right)^{-\left(\ell'+1-i\bar{\Lambda}/\bar{K}\right)} 
   e^{-i\left(\ell'+1-i\bar{\Lambda}/\bar{K}\right)\pi/2}
   \label{E45}
  \end{eqnarray}
 
Because
 \begin{equation}
 \Gamma\left(\ell'+1-i\frac{\bar{\Lambda}}{\bar{K}}\right)=\left| \Gamma\left(\ell'+1-i\frac{\bar{\Lambda}}{\bar{K}}\right)\right|e^{i\delta'}, \quad  \delta_{\ell}'=arg\left(\Gamma\left(\ell'+1-i\frac{\bar{\Lambda}}{\bar{K}}\right)\right)
 \label{E46}
 \end{equation}
 and 
 \begin{equation}
 \Gamma\left(\ell'+1+i\frac{\bar{\Lambda}}{\bar{K}}\right)=\left| \Gamma\left(\ell'+1-i\frac{\bar{\Lambda}}{\bar{K}}\right)\right|e^{-i\delta'} 
 \label{E47}
 \end{equation}
 where  $\delta'$ is a real number. Eq. (\ref{E45}) becomes
 \begin{eqnarray}
 &{}_1F_1\left(\ell'+1-i\frac{\bar{\Lambda}}{\bar{K}},2\ell'+2,-2i\bar{K}r\right)\nonumber  = \frac{\Gamma(2\ell'+2)}{\Gamma(\ell'+1-i\frac{\bar{\Lambda}}{\bar{K}})}\left(\frac{e^{-\frac{\bar{\Lambda}\pi}{2\bar{K}}}\cdot e^{-i\bar{K}r} }{\left(2\bar{K}r\right)^{\ell'+1}}\right) \nonumber\\ &\times 2\sin\left(\bar{K}r+\bar{\Lambda}\ln(2\bar{K}r)/\bar{K}+\delta'-\ell'\pi/2-\pi/2\right)
 \label{E48}
 \end{eqnarray}
   
   By putting Eq. (\ref{E48}) into Eq. (\ref{E42}), we get,
     \begin{equation}
     U_{n,\ell'}(r\rightarrow \infty)=A\cdot \frac{\Gamma(2\ell'+2)}{\Gamma(\ell'+1-i\frac{\bar{\Lambda}}{\bar{K}})}\left(\frac{e^{-\frac{\bar{\Lambda}\pi}{2\bar{K}}} }{\left(2\right)^{\ell'+1}}\right)  2\sin\left(\bar{K}r+\bar{\Lambda}\ln(2\bar{K}r)/\bar{K}+\delta'-\ell'\pi/2\right)
    \label{E49}
     \end{equation}
     On the other hand, using the  asymptotic behavior
      \begin{equation}
      U_{n,\ell}(r\rightarrow \infty)= 2\sin\left(\bar{K}r+\bar{\Lambda}\ln(2\bar{K}r)/\bar{K}+\delta_{\ell}-\ell\pi/2\right)
      \label{E50}
      \end{equation}
     and comparing the arguments of sine term in Eqs. \eqref{E49} and \eqref{E50}, one can then derive the phase shifts,
     \begin{equation}
     \delta_{\ell}=\frac{\pi(\ell-\ell')}{2}+\delta_{\ell}', \
     \label{E51}
     \end{equation}
     
     where $\ell'$ is given by Eq. (\ref{E35}), and the normalization constant also can be evaluated by comparison of coefficients of sine term in Eqs. \eqref{E49} and \eqref{E50} as 
     \begin{equation}
     A=\frac{\Gamma\left(\ell'+1-i\frac{\bar{\Lambda}}{\bar{K}}\right)}{\Gamma(2\ell'+2)}2^{(\ell'+1)}e^{\frac{\pi\bar{\Lambda}}{2\bar{K}}}.
     \label{E52}
     \end{equation}
     \section{ Conclusion}
     In this paper, we have considered  the time-independent Schr\"{o}dinger equation  with a position-dependent effective mass in non central potential. Using the potential form proposed in \cite{b26}, we separated  the deformed Schr\"{o}dinger equation in all coordinates. The radial solution is then obtained, as well as the exact analytical angular solution. We have also studied the scattering states of the deformed Schr\"{o}dinger equation under a non central effective potential and derived the energy eigenvalues and the normalization constant of the radial wave functions, as well as the scattering phase shifts.
     \section*{ Acknowledgment}
      The authors, to be great pleasure in thanking the referee for his/her helpful comments.

\end{document}